\documentclass{article}
\usepackage{flushend}
\usepackage{spconf,amsmath,graphicx}
\usepackage{subfigure}
\usepackage{multirow}
\usepackage{longtable}
\usepackage{amsthm}
\usepackage{amssymb }
\usepackage{amsmath}
\usepackage{bm}
\usepackage{mathrsfs}
\usepackage{amsfonts}
\usepackage{longtable}
\usepackage{multirow}
\usepackage{algorithm, algpseudocode}
\usepackage{cite}
\usepackage{amsmath}

\algnewcommand\Input{\item[\hspace{6pt}\textbf{Input:}]}
\algnewcommand\Output{\item[\hspace{6pt}\textbf{Output:}]}
\algnewcommand\OutputVal{\textbf{output} }
\usepackage{tabularx}
\newcolumntype{L}{>{\raggedright\arraybackslash}X}
\usepackage{makecell}
\usepackage{color}
\usepackage{xcolor}

\title{Data Shapley Value for Handling Noisy Labels: An application in Screening COVID-19 Pneumonia from Chest CT Scans}
\name{\makecell{Nastaran Enshaei$^\dagger$, Moezedin Javad Rafiee, MD$^\ddagger$,
Arash Mohammadi$^\dagger$, and Farnoosh Naderkhani$^\dagger$}}
\address{$~^\dagger$Concordia Institute for Information Systems Engineering, Concordia University, Montreal, Canada\\
$~^\ddagger$Department of Medicine and Diagnostic Radiology, McGill University, Montreal, QC, Canada\\
$~^5$Department of Electrical and Computer Engineering, University of Toronto, Toronto, Canada}
\begin{document}
\ninept
\maketitle
\begin{abstract}
A long-standing challenge of deep learning models involves how to handle noisy labels, especially in applications where human lives are at stake. Adoption of the data Shapley Value (SV), a cooperative game theoretical approach, is an intelligent valuation solution to tackle the issue of noisy labels. Data SV can be used together with a learning model and an evaluation metric to validate each training point's contribution to the model's performance. The SV of a data point, however, is not unique and depends on the learning model, the evaluation metric, and other data points collaborating in the training game. However, effects of utilizing different evaluation metrics for computation of the SV, detecting the noisy labels, and measuring the data points' importance has not yet been thoroughly investigated. In this context, we performed a series of comparative analyses to assess SV's capabilities to detect noisy input labels when measured by different evaluation metrics. Our experiments on COVID-19-infected of CT images illustrate that although the data SV can effectively identify noisy labels, adoption of different evaluation metric can significantly influence its ability to identify noisy labels from different data classes. Specifically, we demonstrate that the SV greatly depends on the associated evaluation metric. 
\end{abstract}

\vspace{.1in}
\section{Introduction} \label{sec:Introduction}
Deep learning has proven remarkable success in several medical fields, including medical imaging, where advanced computer vision algorithms have delivered near human-level performance in specific tasks. Several factors may affect the success of a deep learning algorithm, including model structure, initialization of the model parameters, training methods, and computational hardware. However, the most influential factor is having access to large-scale data sets with reliable labels. 
Collecting large-scale data sets with reliable labels is, however, a significantly challenging task in several applications, particularly within the medical domain. On the one hand, patients' privacy preservation and data sharing protocols prohibit hospitals and clinical institutions from releasing their in-house data sets. On the other hand, labeling medical images requires knowledge and expertise from radiologists and physicians, making the labeling process resource-intensive. To overcome this challenge, extracting pathology labels of large-scale medical images from radiology reports with the help of text mining~\cite{wang2017chestx}, human-machine collaborative techniques~\cite{zhang2020clinically,degerli2021covid}, and the use of non-expert annotators are some of the proposed solutions in the literature~\cite{wang2020noise}. However, the provided labels by such methods are noisy and inaccurate compared to manually labeled images~\cite{oakden2020exploring}. Consequently, handling noisy labels in medical imaging data sets is crucial for developing high-performance processing/learning models.

\noindent
\textbf{Prior work:}
Learning from a training data set with noisy labels has been a challenging task within the deep learning domain for a long time. Research studies indicate that noisy labels may have a more significant adverse effect on the performance of deep learning models than the noises in data attributes/measurements~\cite{zhu2004class}. Proposed techniques for handling noisy labels in medical imaging data sets include weakly supervised learning~\cite{hu2020weakly,wang2020weakly,laradji2021weakly}, customized training methods~\cite{wang2020noise}, and re-weighting training samples~\cite{xue2019robust}. Another approach to tackle the noisy label problem is to adopt Shapley Value (SV), a cooperative game theoretical method, where each training point is considered a player in the training game, and its contribution to any subset of players is measured using a performance evaluation metric. Researchers have leveraged the SV for data valuation and measuring the quality of data points in different deep learning applications~\cite{maleki2013bounding,lundberg2017unified,merrick2020explanation,song2016shapley,vstrumbelj2014explaining,ghorbani2019data,tang2021data,jia2019towards}. In~\cite{wang2020principled} and~\cite{khuri2021value}, authors utilized the SV in a federated learning study to measure the quality of data from each participator. A recent study~\cite{tang2021data} has investigated the capability of the SV in quantifying the importance of training data points in the performance of the learning model in a large-scale X-ray data set. 
It should be noted that the SV of a data point is not a unique value and depends on the learning model, the evaluation metric, and other data points collaborating in the training game. So far, research studies have used only one specific evaluation metric in their SV computation process, such as accuracy metric in~\cite{wang2020principled,tang2021data} and Area under the ROC curve (AUC-ROC) metric in~\cite{khuri2021value} to demonstrate the data SV capability in measuring the quality of data points. However, the effect of utilizing different evaluation metrics in computing the SV, detecting the noisy labels, and measuring the data points' importance in the training process has never been investigated. 

\noindent
\textbf{Contributions:} This study investigates effects of incorporating different evaluation metrics in determining the data SV and quantifying the importance of each training point in the model's  performance. To the best of our knowledge, this is the first study that explores adoption of various evaluation metrics in measuring the data SV and detecting noisy input labels. The data SV measures the contribution of each training point to all possible subsets of the training set, having an exponential complexity in the size of the training set. Therefore, we use a permutation sampling approach to approximate the data SV. We assess the capability of data SV obtained by commonly-used classification metrics, including accuracy, recall, and specificity in detecting the noisy labels through a set of experiments with different noise levels. 
The experiments are performed based on a COVID- 19 screening task from chest CT scans by implementing a lightweight Capsule-network-based classifier~\cite{Hinton2018} to extract discriminative features from chest CT scans and distinguish COVID-19-infected CT images from normal ones~\cite{Mohammadi2021}. The Capsule-network-based classifier can represent each CT image via a small feature vector. Since calculating the SV requires retraining the learning model on multiples coalitions of training samples, we use a fast classifier to make our data valuation process time-efficient. Therefore, the extracted feature vector is fed into a logistic regression classifier for final decision making. The results indicate that the measured data SV is not unique and is highly dependent on the evaluation metric. We demonstrate that despite the great potential of the data SV in detecting noisy labels, its performance is extremely affected by the adopted evaluation metric. 

\section{Data Shapley Value} \label{sec:method}
\begin{figure*}[t!]
\centering
\includegraphics[width=0.9\textwidth]{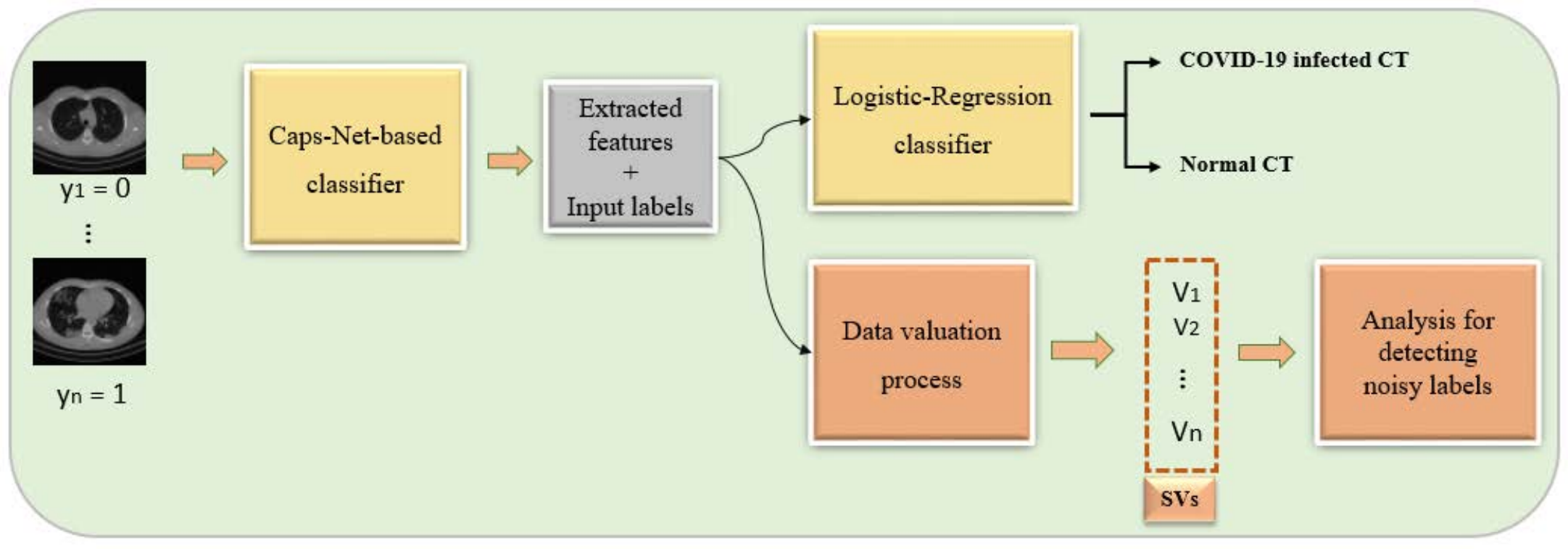}
\caption{\footnotesize The proposed framework to quantify the quality of data and discriminate COVID-19 infected CT images from normal ones. \label{fig:pipe}}
\end{figure*}

This study implements the data SV in its data valuation approach for identifying training samples with noisy labels. Data SV~\cite{shapley201617} is a well-known measure in cooperative game theory for assigning a fair payoff to each player of the game by taking into account its contribution to all possible coalitions of the players. The training process of an ML/DL model can be assumed as a game where training data points are players that collaborate to achieve the highest model performance. Therefore, the SV can measure the importance of training samples in the performance of the ML/DL model~\cite{maleki2013bounding,lundberg2017unified,merrick2020explanation,song2016shapley,vstrumbelj2014explaining,ghorbani2019data,tang2021data,jia2019towards}. Having a training set $\textit{D}$, the SV of a data point $i$ $\epsilon$ $\textit{D}$ is calculated as 
\begin{equation} \label{eq:SV_eq}
\text{$SV_i$} = \frac{1}{N} \sum_{S \subseteq D - \{ i \}}\frac{1}{\binom{N - 1}{|S|}} [V \{ S \cup i \} - V \{ S \}],
\end{equation}
where $S$ is any possible subset of $D$ not including $i$, $N$ is the size of the training set $D$, and $V$ is an evaluation metric for measuring the model performance. $V \{ S \}$ is the performance of the model trained on the subset $S$ and measured by the evaluation metric $V$. In other words, the SV of a training data point $i$ is its average marginal contribution to all subsets of $S$, which can be interpreted as a quality measure for data assessment. However, computing SV for a player requires computing its marginal contribution to all possible coalitions of game players, which has exponential complexity in the number of players. Moreover, in data SV computation, calculating each marginal contribution, $V \{ S \}$ requires training the model on S that makes the computation process more time-consuming. To overcome this challenge, we use a permutation sampling method~\cite{jia2019towards} to approximate the data SV. For this purpose, we use an equivalent form of Eq.~\ref{eq:SV_eq} as follows

\begin{equation} \label{eq:SV2_eq}
\text{$SV_i$} = \frac{1}{N!} \sum_{\pi \epsilon \Pi} [V \{ S_i^\pi \cup i \} - V \{ S_i^\pi \}],
\end{equation}
where $\Pi$ is all possible permutations of data points, $\pi \epsilon \Pi$ is a sample permutation of data points, and $S_i^\pi$ is the set of training points coming before data point $i$ in permutation $\pi$. Indeed, in each random permutation, we can assume that data points are joining the training game in random order. Then, each data point will receive the marginal contribution that his collaboration brings to the training points which have already played their role in the game. The data SV for each training point would be its average marginal contributions over all possible permutations. In practice, using a permutation sampling approach, the data SV will converge after $3N$ permutations~\cite{ghorbani2019data}.
It is worth mentioning that the SV of a data point is not a unique value and depends on the learning model, the evaluation metric, and other data points collaborating in the training game. So far, other studies have used a specific evaluation metric in their SV computation process and demonstrated the data SV capability in measuring the quality of data points. Here, we perform a set of experiments to investigate the effect of different evaluation metrics in measuring the SV, quantifying the quality of training points, and detecting noisy labels. We use the standard evaluation metrics in classification tasks, including accuracy, recall, and specificity, and discuss their effectiveness in detecting noisy labels in both classes of data. Our data valuation process, illustrated in Fig.~\ref{fig:pipe}, includes: i) training a deep classifier on a training set, ii) extracting high-resolution features from training CT images using the trained deep classifier and feeding them into a fast classifier, which is logistic regression in this study, iii) calculating the SV of each training point based on different evaluation metric, iv) analyzing the obtained SVs and detecting noisy labels. 

The deep classifier used for discriminating COVID-19 infected CT images from the normal ones is a lightweight DL model containing two convolutional and two Capsule layers. A batch-normalization layer follows the first convolutional layer, and a max-pooling layer follows the second one. The ReLU activation function is applied after each convolutional layer to capture non-linear patterns. The number of channels for each convolutional layer is $64$. The output of the max-pooling layer is reshaped and fed to a Capsule layer to extract high-resolution features from CT images. Finally, the last Capsule layer predicts the probability that each CT image belongs to infected or non-infected classes. It is noteworthy that screening COVID-19-infected CT images on a slice-level basis can be a primary step in detecting COVID-19 patients from healthy cases. We use a weighted loss function to tackle the imbalance dataset, considering a higher penalty to the samples from minority class, which is COVID-19 infected slices in our case. By extracting high-resolution features from the last Capsule layer, each CT image with a matrix size of $512 \times 512$ is presented in a $1 \times 16$ vector. Next, extracted features from the training set and their corresponding labels are used in the SV computation process. For more information about Capsule network-based classifiers and the weighted loss function, please refer to Reference~\cite{heidarian2021covid}. 
 
\section{Dataset description} \label{sec:data}
\begin{table}[t!]
\small
\centering
\caption{\footnotesize Training, validation and test sets used in our experiments.\label{tab:datasplit}}
\begin{tabular}{|c|c|c|c|}
\hline
\textbf{} & \textbf{COVID-19} & \textbf{Normal} & \textbf{Total} \\[1ex]
\hline
Training set & $100$ & $400$ & $500$ \\
\hline
Validation Set & $50$ & $200$ & $250$ \\
\hline
Test Set & $241$ & $759$ & $1000$ \\
\hline
\end{tabular}
\end{table}
The dataset used in our experiments is a combination of two different open-access datasets. For non-COVID-19 CT images, we used a subset of $50$ normal cases from our recently released dataset referred to as the ``COVID-CT-MD''~\cite{Afshar2020a}, which is available through Figshare~\footnote{https://figshare.com/s/c20215f3d42c98f09ad0}. For CT images with the evidence of COVID-19 lesions, we used a public dataset containing chest CT volumes of 10 COVID-19 patients~\cite{ma2020towards}, where three expert radiologists have annotated COVID-19 manifestations. In both datasets, the matrix size of the CT images is $512\times512$ pixels. 
We randomly split the dataset into three independent groups, as presented in table~\ref{tab:datasplit}, including $100$, $50$, and $241$ COVID-19 infected CT scans, and $400$, $200$, and $759$ normal CT scans for training, validation, and test sets. The extracted features from the training set are used to train the logistic regression model over multiple permutation sampling, and the test set is used for measuring the model performance during the SV computation process. 

\section{Experimental Results} \label{sec:result}
\begin{figure}[t!]
\centering
\includegraphics[width=0.48\textwidth]{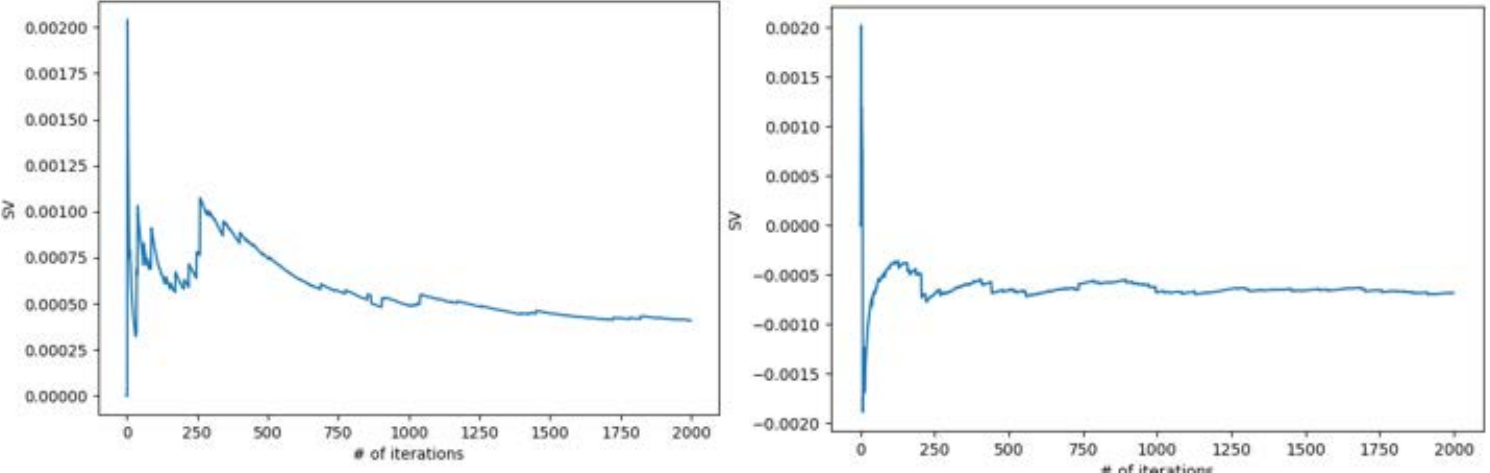}
\caption{\footnotesize The convergence of SV for randomly selected training samples. The x-axis indicates the number of iteration in SV computation process and the y-axis indicates the estimate of SV.  \label{fig:convergence}}
\end{figure}
As the pre-processing step, each CT slice is normalized based on its pixel intensities' mean and standard deviation. Furthermore, we utilize the real-time data augmentation method to enhance the model's performance on unseen data. Each mini-batch of original images during the training process is transformed into synthetic images using conventional data augmentation strategies such as zooming, shifting, and shearing. Over the training' epochs, the model will observe each augmented image only once, resulting in an improvement in the model's generalization. In each mini-batch, 16 CT images are fed to the network. The Adam optimization algorithm with an initial learning rate of $0.001$ minimizes the loss function over the training process. The number of passes through the training set is set to $100$. However, to mitigate the model over-fitting, the training process will stop whenever the loss function on the validation set is not reduced over five epochs. Finally, the SV computation is performed using the training set as the players for training the logistic regression classifier and the test set for measuring the model performance trained on any possible subset of training points. As mentioned previously, we have $100$ COVID-19 infected CT slices and $400$ normal CT images as our training set. Since we aim to deal with noisy labels in the training set, we manually add some noises in the labels of our training set and investigate the capability of the SV obtained based on different evaluation metrics in detecting noisy labels. We run three experiments with three different noise levels, including $10\%$, $20\%$, and $30\%$ in each class of data. Therefore, 10, 20, and 30 positive CT images and 40, 80, and 120 normal CT scans have incorrect labels in experiment I, II, and III, respectively.  

\begin{table*}[t!]
    \centering
    \caption{\footnotesize The efficiency of data SV obtained based on different evaluation metrics in detecting noisy labels of each class of data in Exp $I$, $II$, and $III$ with the noise level of $10\%$, $20\%$, and $30\%$.}
    \includegraphics[scale=0.7]{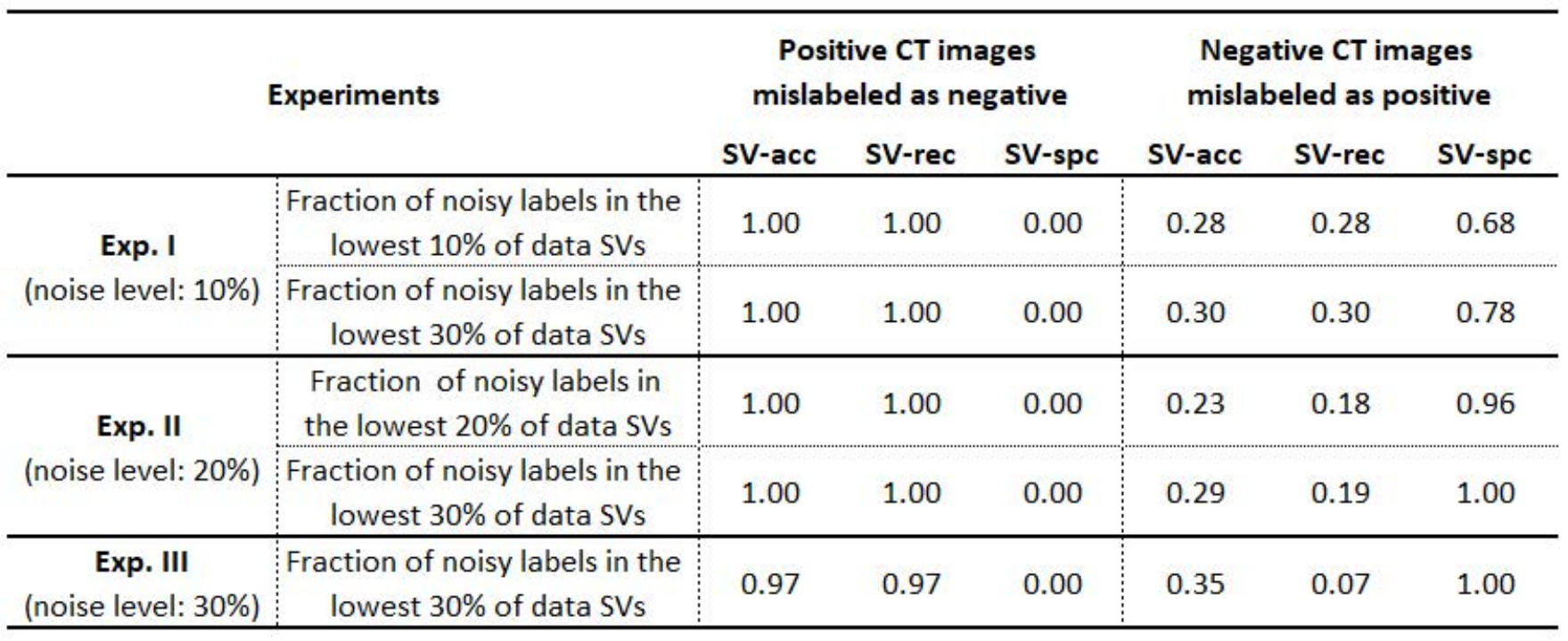}
    \label{tab:SVres}
    \vspace{-.1in}
\end{table*}
\begin{figure}[t!]
\centering
\includegraphics[width=0.49\textwidth]{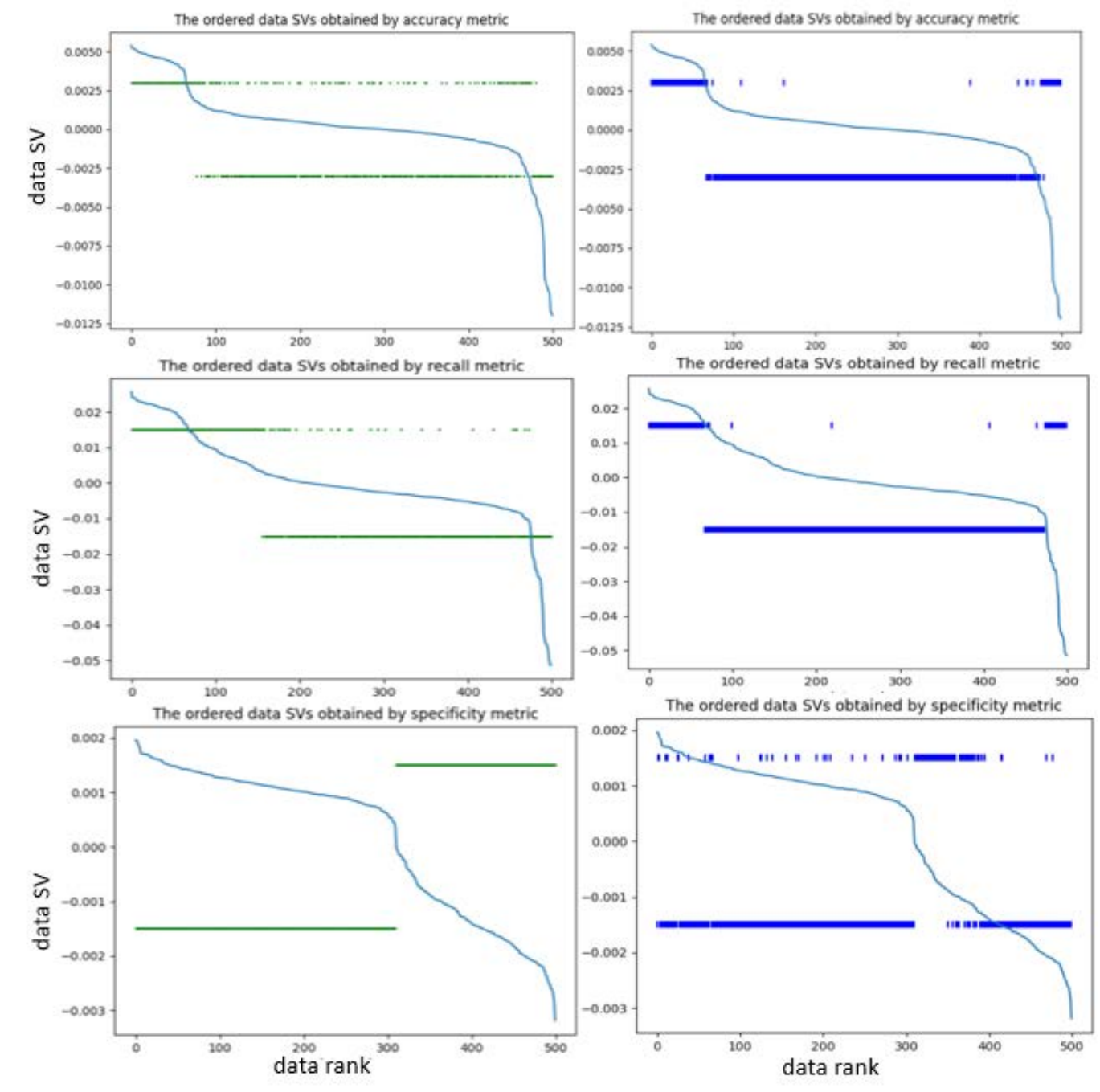}
\caption{\footnotesize Mapping between the data SV and the type of data class derived via different evaluation metrics (Plots are related to the Exp. III).  \label{fig:SV-classes}}
\end{figure}
%
In each experiment, first, we train the Capsule-network-based classifier using the training set with noisy labels. Next, the extracted features from the trained classifier are fed into a logistic regression classifier to compute the data SV. It is worth mentioning that computing the SV for each training point requires re-training the classifier on multiple subsets of training points, making the data evaluation process time-intensive. Hence, we use a fast classifier such as logistic regression to make the computation process affordable. We adopt different evaluation metrics, including accuracy, recall, and specificity, to calculate the data SV. We run the computation process for $2000$ permutations to assure the convergence of the estimated data SVs. Fig.~\ref{fig:convergence} presents the convergence of the SV for two randomly selected training samples. As can be observed, the estimated SV converges after $3N = 1500$ permutations, which is in agreement with previous works~\cite{ghorbani2019data}.

Next, we order the training points based on their estimated SV and mapped them with the corresponding label in our input labels that contains some noisy labels (green points in column left) and the ground-truth labels, which are correct labels (blue lines in right column), as visualized in Fig.~\ref{fig:SV-classes}. As can be observed, when adopting accuracy or recall metrics, all training points with the highest data SV belong to the positive class. In contrast, the specificity metric ranks all training points with the negative input labels before the data points with positive input labels. The plots also indicate that although the specificity metric puts all images with the positive input label among the lowest data SVs, the least valuable data points are the ones mislabeled as positive classes (their ground-truth label is negative). This reveals the capability of the SV in identifying noisy labels with the negative ground-truth labels (mislabeled as the positive images in the input label) when adopting the specificity metric. On the contrary, the majority of the least valuable training points obtained by the accuracy and recall metrics belong to negative input labels. However, according to the plots, the accuracy metric assigns more value to the images with negative input labels than recall metric. In addition, by comparing the plots in left and right columns, we can conclude that the least valuable data points obtained by both accuracy and recall metrics are the images mislabeled as the negative ones (have positive ground-truth labels). This indicates the possibility of detecting noisy labels with the positive ground-truth labels when using the accuracy or recall metrics in computing the SV.       

Furthermore, we investigate the effect of utilizing different evaluation metrics in the SV computation process for identifying noisy labels. We determined the fraction of noisy labels ranked as the $10\%$, $20\%$, and $30\%$ of the lowest data SVs for experiments $I$, $II$, and $III$, respectively. We also consider the fraction of noisy labels ranked as the lowest $30\%$ data SVs in all experiments. As demonstrated in Table~\ref{tab:SVres}, when adopting accuracy or recall metrics, all noisy labels with the positive ground-truth label (which had been incorrectly labeled as negative ones in input labels) have been correctly identified in experiments $I$ and $II$. Conversely, when utilizing specificity metric in the SV computation process, no noisy label with the positive ground-truth label (mislabeled as negative images in the input labels) is ranked between the lowest $30\%$ of data SVs. The results indicate that the adoption of the specificity metric has been more successful in detecting noisy labels with the negative ground-truth label (incorrectly labeled as positive ones in input labels). All noisy labels with the negative ground-truth labels (both correctly labeled and mislabeled ones) have been ranked in the lowest $30\%$ data SVs. Although the accuracy and recall metrics show the same performance in detecting noisy labels with a negative ground-truth label, the performance of the accuracy metric succeeds the recall metric as the noise level increases. 

It should be mentioned that, to the best of our knowledge, the adoption of different evaluation metrics in measuring the data SV has not been discussed in previous researches. However, Reference~\cite{tang2021data}, which has performed an SV-based data valuation study using the accuracy metric on an X-ray data set containing $2000$ training points, illustrates that all the $100$ training points with the highest SVs belong to the positive class. Besides, with the help of three radiologists, they figured out that out of the $100$ lowest data SVs in their experiment, there were $65$ mislabeled images where $80\%$ of them were positive images that had been incorrectly labeled as negative ones. Indeed, their experiments are in accordance with our results, confirming that by utilizing the accuracy metric in computing the SV, the images with positive labels will receive the highest values. In addition, it would be more likely to detect mislabeled images with ground-truth positive labels (mislabeled as negative ones). Our findings show that, while the data SV has a lot of potential for detecting noisy labels in training sets, it is extremely dependent on the evaluation metric used.     

\section{Conclusion}  \label{sec:con}
This research explores the effect of using different evaluation metrics in data SV computation and its capability and limitation in identifying noisy labels in a training set. We examine the standard evaluation metrics in classification tasks, including accuracy, recall, and specificity in calculating the data SV on a chest CT scan data set. Our findings show that, while the data SV has a lot of potential for detecting noisy labels in training sets, it depends highly on the evaluation metric used. We also demonstrate that the data SV is not a unique value and will differ by incorporating different evaluation metrics. This research study conducted experiments on a binary classification task and a limited data set. We leave the implementation of data SV based on various evaluation metrics and investigating its capabilities and limitations in detecting noisy labels in a multi-institutional data set and a multi-class classification problem to the future. Exploring the dependency of the data SV on different types of learning algorithms is another future direction of the present work.

\vspace{-.1in}

\bibliographystyle{ieeetr}
\footnotesize
\bibliography{refs}

\end{document}